\documentclass{PoS}

\def \as {\relax\ifmmode\alpha_s\else{$\alpha_s${ }}\fi}
\def\e{\epsilon}

\title{All-order results for infrared and collinear singularities in massless gauge theories}

\ShortTitle{All-order results for infrared and collinear singularities}

\author{Lance J. Dixon%
\thanks{Research supported by the US Department of 
        Energy under contract DE--AC02--76SF00515.}\\ 
        SLAC National Accelerator Laboratory, Stanford University\\
        E-mail: \email{lance@slac.stanford.edu}}

\author{Einan Gardi\\
        School of Physics, The University of Edinburgh\\
        E-mail: \email{einan.gardi@gmail.com}}

\author{\speaker{Lorenzo Magnea}%
         \thanks{Work supported in part by the European Community Network `HEPTOOLS', contract MRTN-CT-2006-035505.} 
         \\
       CERN, PH Department, TH Unit; \\
       Dipartimento di Fisica Teorica, Universit\`a di Torino, and
       INFN Sezione di Torino \\
       E-mail: \email{magnea@to.infn.it}}

\abstract{We review recent results concerning the all-order structure of 
infrared and collinear divergences in massless gauge theory amplitudes. 
While the exponentiation of these divergences for nonabelian gauge 
theories has been understood for a long time, in the past couple of 
years we have begun to unravel the all-order structure of the anomalous 
dimensions that build up the perturbative exponent. In the 
large-$N_c$ limit, all infrared and collinear divergences are determined 
by just three functions; one of them, the cusp anomalous dimension, 
plays a key role also for non-planar contributions. Indeed, all infrared 
and collinear divergences of massless gauge theory amplitudes with any 
number of hard partons may be captured by a surprisingly simple 
expression constructed as a sum over color dipoles. Potential
corrections to this expression, correlating four or more hard partons 
at three loops or beyond, are tightly constrained and are currently 
under study.} 

\FullConference{RADCOR 2009 - 9th International Symposium on 
Radiative Corrections (Applications of Quantum Field Theory to 
Phenomenology), October 25-30 2009, Ascona, Switzerland.\\
{\tt SLAC--PUB--13938, Edinburgh 02/2010, CERN-PH-TH/2010-017,
DFTT 01/2010.}}

\begin{document}

\section{Introduction}
\label{intro}

The long-distance behavior of gauge theory amplitudes and cross sections has been the subject of theoretical studies for nearly three quarters of a century. After such a long time, one might imagine 
that further progress, if any, should be slow and incremental: on 
the contrary, the past few years have witnessed several new  
developments, some of them surprising. This has been due in part 
to the phenomenological requirements set by the beginning of LHC
operation, which places unprecedented pressure on theorists to come 
up with precise and reliable predictions for very complex QCD processes.
On the other hand, `pure theory' has also played a role: for example, 
insights have come from the study of the maximally supersymmetric 
${\cal N} = 4$ Yang-Mills theory; also, part of the general 
motivation for these studies remains the fact that long-distance 
singularities provide a gateway from perturbative calculations to the 
non-perturbative content of the theory.

Much of the knowledge accumulated in earlier work on the structure 
of singularities for gauge theory $S$-matrix elements, which forms the 
basis for more recent developments,  can be summarized in a single
formula\footnote{For reviews of the methods leading to Eq.~(\ref{facto}) see, for example,~\cite{reviews}.}, expressing the factorization 
of fixed-angle amplitudes into separate functions, 
responsible for infrared poles, collinear poles, and finite remainders. 
Choosing a basis in the space of available color structures for a given 
$n$-particle amplitude, by picking suitable color tensors $c_L^{a_1,
\ldots, a_n}$, the amplitude $M^{a_1, \ldots, a_n}$ can be 
expressed in terms of its components $M_L$. They obey
\begin{eqnarray}
  M_{L} \left(p_i/\mu, \as (\mu^2), \e \right) & = & 
  {\cal S}_{L K} \left(\beta_i \cdot \beta_j, \as (\mu^2), \e \right) \,  
  H_{K} \left( \frac{p_i \cdot p_j}{\mu^2},
  \frac{(p_i \cdot n_i)^2}{n_i^2 \mu^2}, \as (\mu^2), \e \right)
  \nonumber \\ && \hspace{-10mm} \times \, \,
  \prod_{i = 1}^n \left[ {\displaystyle J_i 
  \left(\frac{(p_i \cdot n_i)^2}{n_i^2 \mu^2},
  \as (\mu^2), \e \right)} \Bigg/ {\displaystyle {\cal J}_i 
  \left(\frac{(\beta_i \cdot n_i)^2}{n_i^2}, \as (\mu^2), \e \right)} \, \right] 
  \,\, ,
\label{facto}
\end{eqnarray}
where $p_i$, $i = 1, \ldots, n$, are the external hard momenta, which
are assumed to form invariants $p_i \cdot p_j$ of a common parametric 
size $Q^2$. In Eq.~(\ref{facto}), the functions $J_i$ collect all collinear
singularities associated with virtual gluons emitted in direction of parton
$i$. In order to factorize these singularities, it is necessary to introduce
`factorization vectors' $n_i^\mu$, $n_i^2 \neq 0$, which play a 
threefold role: first, they ensure gauge invariance of the operator matrix 
element defining $J_i$, which includes an infinite Wilson line in the 
direction $n_i$; these Wilson lines act as absorbers, replacing the other hard partons in the amplitude, and collecting the gluons emitted by 
parton $i$ without generating extra singularities; finally, $n_i$ can be 
physically interpreted as a vector separating gluons which are collinear
to $p_i$ (those whose momenta $k$ satisfy $k \cdot p_i < k \cdot 
n_i$) from (soft) gluons emitted at large angles; in this sense $n_i$
can be properly interpreted as a factorization vector. Note that the 
jet functions $J_i$ are color singlets: infrared (soft) singularities, on 
the other hand, are not color diagonal, and therefore they are 
organized in a matrix, ${\cal S}_{L K}$, which is purely eikonal. 
It is defined in terms of a product of Wilson lines characterized 
by the directions and color representations of the hard partons. 
Being purely eikonal, ${\cal S}_{LK}$ can only depend on the 
velocities $\beta_i$ of the hard partons, defined by dividing 
out the common hard scale $Q$, taking $p_i \propto Q \beta_i$. 
With these definitions, gluons that are both soft and collinear to one
of the hard partons have been counted twice; it is however simple 
to subtract the double counting: one just needs to divide each jet
$J_i$ by its own eikonal approximation, denoted by ${\cal J}_i$ in 
Eq.~(\ref{facto}). The vector of hard functions $H_K$, finally, collects
all finite remainders, and each component is finite as $\e \to 0$. Note 
that the matrix element $M$ has been normalized to be dimensionless:
the functional dependences of the various factors in Eq.~(\ref{facto})
reflect this fact, and will be further discussed below. In the following,
we will briefly outline the consequences of the factorization presented
in Eq.~(\ref{facto}), beginning with the simpler case of
amplitudes at large $N_c$.

\section{Form factors and large-$N_c$ amplitudes}
\label{dms}

The simplest instance of Eq.~(\ref{facto}) is given by parton form 
factors, which describe the scattering of a parton by an electroweak 
current\footnote{We thank George Sterman, who co-authored 
Ref.~\cite{Dixon:2008gr}, where the results of this section were 
derived.}.  This is  a color-singlet process, so that the soft matrix
${\cal S}$ is just a single function. One may then derive an evolution
equation by imposing that the form factor be independent of the factorization vectors $n_i$. The solution to this equation is especially 
simple and transparent in dimensional regularization~\cite{Magnea:1990zb}: within this framework, one may take advantage of the fact 
that the $d$-dimensional running coupling vanishes as a power of the 
scale for $d > 4$, in order to impose as a boundary condition 
that radiative corrections should vanish at $Q^2 = 0$. 
One finds that the form factor exponentiates 
as~\cite{Dixon:2008gr}
\begin{equation}
  \Gamma \left( Q^2, \e \right) \, = \, 
  \exp \left\{ \frac{1}{2} 
  \int_0^{- Q^2} \frac{d \xi^2}{\xi^2} \left[
  G \Big(\overline{\alpha}  \left(\xi^2, \e \right), \e \Big)  -
  \frac{1}{2} \, \gamma_K \Big( \overline{\alpha}
  \left(\xi^2, \e \right) \Big) \, \log \left(\frac{- Q^2}{\xi^2} \right) \right] 
  \right\} \, .
\label{formfacto}
\end{equation}
The first remarkable fact about Eq.~(\ref{formfacto}) is that all 
singularities are generated by integrating over the scale of the 
$d$-dimensional coupling $\overline{\alpha}$: the functions 
$G$ and $\gamma_K$ are finite as $\epsilon \to 0$ and universal.
Specifically, $\gamma_K (\as)$ is the cusp anomalous 
dimension~\cite{Korchemsky:1985xj}, governing 
ultraviolet singularities for correlators of pairs of light-like Wilson 
lines originating at a cusp, and responsible in this case for 
double (infrared-collinear) poles; it depends only on the color
representation of the hard parton. The function $G$ generates 
single poles, as well as finite contributions; it can be expressed 
in terms of operator matrix elements involving Wilson lines as well as
elementary fields~\cite{Dixon:2008gr}, so it also depends on the
the spin of the hard parton. Perhaps more interestingly, it can be 
decomposed as
\begin{equation}
  G (\as, \e) = 
   2 \, B_\delta \left( \as \right)  + G_{\rm eik} \left( \as \right) + 
   \beta(\as) \, \partial E_H \left( \as, \e \right)/\partial \as~.
\label{decompo}
\end{equation}
In Eq.~(\ref{decompo}), $B_\delta(\as)$ is the virtual part of the 
diagonal Altarelli-Parisi splitting function, while $G_{\rm eik} (\as)$ 
is a subleading anomalous dimension for Wilson line correlators, 
associated with the eikonal approximation to the form factor; the 
last term generates finite contributions only, and it vanishes in
a conformal theory. We learn that the only non-eikonal contributions 
to the singular behavior of the form factor are contained 
in the collinear function $B_\delta (\as)$. Notably, in the 
conformal case, Eq.~(\ref{decompo}) holds also at strong coupling, 
in the large $N_c$ limit~\cite{Alday:2009zf}.

The second remarkable fact about Eq.~(\ref{formfacto}) is that it
generates all singularities not only for form factors, but for all 
large-$N_c$ multiparton fixed-angle amplitudes as 
well~\cite{Sterman:2002qn,Bern:2005iz}.
In the planar limit, gluons are confined to propagate inside 
`wedges' bounded by neighboring hard partons: the full amplitude 
becomes then a product of (square roots of) form factors, up to finite
corrections.

Finally, the fact that long-distance singularities are completely 
encoded in the running coupling has important consequences for
conformal theories, such as ${\cal N} = 4$ Super-Yang-Mills (SYM). 
In dimensional regularization, the coupling for these theories runs
simply according to $\overline{\alpha}(\mu^2,\e) = \left( 
\mu^2/\mu_0^2 \right)^{-\e} \, \overline{\alpha} (\mu_0^2, 
\e)$, so that all integrations in Eq.~(\ref{formfacto}) are 
trivially performed. This fact was exploited 
in~\cite{Bern:2005iz} to study and test the all-order 
behavior of amplitudes in ${\cal N} = 4$ SYM. By the same token, 
one derives~\cite{Dixon:2008gr} a strikingly simple relation tying
the analytic continuation of the form factor to the cusp anomalous dimension. One finds that in any conformal gauge theory
\begin{equation}
  \left| \frac{\Gamma(Q^2)}{\Gamma(- Q^2)} \right|^2 = \exp \left[ 
  \frac{\pi^2}{4} \, \gamma_K \left( \alpha_s (Q^2) \right) \right]~.
\label{tregam}
\end{equation}
Eq.~(\ref{tregam}) ties together two quantities that are finite (and
actually independent of $Q^2$) in $d = 4$, and are non-perturbatively
defined. It may thus be argued to be an exact result for a class of non-trivial four dimensional gauge theories, which may at some point 
become testable at strong coupling.

\section{Beyond the large-$N_c$ limit}
\label{gm}

In  order to go beyond the large-$N_c$ limit, one must revisit 
the functional dependences in Eq.~(\ref{facto}). Functions defined 
in terms of Wilson lines with velocities $n_i$ (with $n_i^2 
\neq 0$) depend homogeneously on the velocity vectors, reflecting 
the classical invariance of their operator definitions under rescalings 
$n_i^\mu \to \kappa_i n_i^\mu$. This is not the case for functions 
of the light-like Wilson lines with velocities $\beta_i$. The reason can 
be traced to the fact that these functions acquire new collinear 
divergences, which make the classical invariance under 
$\beta_i^\mu \to \kappa_i \beta_i^\mu$ anomalous at the 
quantum level. The anomaly is precisely expressed by the cusp 
anomalous dimension, which governs the superposition of soft 
and collinear singularities. This fact has remarkable 
consequences on the singularity structure. Indeed, one may
construct a reduced soft matrix, where all soft-collinear double poles
are cancelled, so that the anomaly in rescaling invariance is absent.
It is defined by~\cite{Sterman:2002qn,Gardi:2009qi}
\begin{equation}
  \overline{{\cal S}}_{L K} \left(\rho_{i j},\as(\mu^2), \e \right) = 
  \frac{{\cal S}_{L K} \left(\beta_i 
  \cdot \beta_j, \as(\mu^2), \e \right)}{\displaystyle \prod_{i = 1}^n 
  {\cal J}_i \left(\frac{(\beta_i \cdot 
  n_i)^2}{n_i^2}, \as(\mu^2), \e \right)}~.
\label{sbar}
\end{equation}
While the functions ${\cal S}_{LK}$ and ${\cal J}_i$ separately have 
double poles, and can thus depend on variables like $\beta_i \cdot \beta_j$, or $x_i \equiv (\beta_i \cdot n_i)^2/n_i^2$, the reduced
matrix $\overline{\cal S}_{LK}$ has only single poles, and can only 
depend on rescaling-invariant combinations of $\beta_i$ and $n_i$,
which can only be constructed out of the variables $\rho_{ij} \equiv
(\beta_i \cdot \beta_j)^2/x_i x_j$. Since all functions entering 
Eq.~(\ref{sbar}) are multiplicatively renormalizable, this recombination 
must be reflected in their respective anomalous dimensions, which must obey
\begin{equation}
\Gamma^{\overline{{\cal S}}}_{I J} \left(\rho_{i j}, \as(\mu^2) \right) 
\, =  \, \Gamma^{{\cal S}}_{I J} \left( \beta_i \cdot \beta_j, \as(\mu^2), 
\e \right) - \delta_{I J} \sum_{k = 1}^n
\gamma_{{\cal J}_k} \left( x_k, 
\as(\mu^2), \e \right) \, .
\label{gammasbar}
\end{equation}
We see from Eq.~(\ref{gammasbar}) that singular terms in the matrix
$\Gamma^{\cal S}$ must be canceled by those in the eikonal jet 
anomalous dimensions $\gamma_{\cal J}$, and must therefore be 
diagonal. Furthermore, finite diagonal contributions must conspire to 
reconstruct a dependence on $\rho_{ij}$, combining $\beta_i \cdot 
\beta_j$ with $x_i$ and $x_j$. Finally, off-diagonal terms in $\Gamma^{\cal S}$ must be finite, and must by themselves depend only on 
rescaling-invariant combinations of $\beta_i$'s. Such combinations
exist only for amplitudes with at least four-particles, and must be 
built out of conformal cross-ratios of the form $\rho_{ijkl} 
\equiv (\beta_i \cdot \beta_j) (\beta_k \cdot \beta_l)/(\beta_i 
\cdot  \beta_k) (\beta_j \cdot \beta_l)$. These powerful 
constraints can be summarized in a single set of 
equations~\cite{Gardi:2009qi}, linking the matrix 
$\Gamma^{\overline{\cal S}}$ to the cusp 
anomaly, and correlating kinematic and color degrees of freedom 
for any number of partons and at finite $N_c$. They are given by 
\begin{equation}
\sum_{j \neq i} \frac{\partial}{\partial \ln(\rho_{i j})} \,
\Gamma^{{\overline{\cal S}}}_{MN} \left( 
\rho_{i j}, \as \right) = \frac{1}{4} \, \gamma_K^{(i)} 
\left( \as \right) \,\delta_{MN}  \qquad \forall i\,,
\label{constraint}
\end{equation}
where we made explicit the fact that the cusp anomalous dimension
depends on the color representation of the selected parton $i$. An
analogous equation was independently derived, using the methods 
of soft-collinear effective field theory, in~\cite{Becher:2009cu}.

One may now make the further assumption that the cusp anomalous 
dimension depend on the representation only through an overall factor 
of the quadratic Casimir operator, {\it i.e.} $\gamma_K^{(i)} (\as) = 
C_i \widehat{\gamma}_K (\as)$, with $\widehat{\gamma}_K (\as)$ 
a universal function. This assumption (`Casimir scaling') is true up to 
three loops, and arguments were given in~\cite{Becher:2009cu} 
indicating that it should remain valid at four loops. Adopting the 
basis-independent notation of color generators, and thus writing 
$C_i = {\rm T}_i \cdot {\rm T}_i$, one may construct an explicit solution of 
Eq.~(\ref{constraint}). Indeed, the sum-over-dipoles formula
\begin{equation}
\Gamma^{\overline{\cal S}}_{\rm dip}
\left(\rho_{i j}, \as \right) \, = \, - \frac18 \,
\widehat{\gamma}_K\left(\as \right) \, \sum_{j \neq i} \,
\ln(\rho_{ij})  \, \, \mathrm{T}_i \cdot \mathrm{T}_j  \, + \, 
\frac12 \, \widehat{\delta}_{{\overline{\cal S}}} ( \as )  
\sum_i \mathrm{T}_i \cdot \mathrm{T}_i
\label{minsol}
\end{equation}
satisfies Eq.~(\ref{constraint}), as is easily shown using color 
conservation, $\sum_i {\rm T}_i = 0$. If desired, the residual anomalous dimension $\widehat{\delta}_{{\overline{\cal S}}} (\as)$ can be 
recombined with the color singlet contributions arising from jet 
functions in Eq.~(\ref{facto}): a sum-over-dipoles formula then
organizes all infrared and collinear divergences of fixed-angle amplitudes, 
for an arbitrary number of partons~\cite{Becher:2009cu,Gardi:2009zv}.

\section{Beyond the dipole formula?}
\label{dgm}

Eq.~(\ref{minsol}) is remarkable, as it implies that color correlations
induced by soft gluons are drastically simpler than expected on the 
basis of a diagrammatic analysis. One must ask what corrections, if 
any, might  arise, that would be compatible with the constraint equation 
(\ref{constraint}). Clearly, one possibility is a breakdown of Casimir 
scaling, {\it i.e.} the presence of higher-rank Casimir operators in the
cusp anomalous dimension at sufficiently high order. The 
only other possibility allowed by Eq.~(\ref{constraint}) is the addition
of a solution to the associated homogeneous equation.
One may write in full generality
\begin{equation}
\Gamma^{{\overline{\cal S}}}\left(\rho_{i j}, \alpha_s \right)
= \Gamma_{\rm dip}^{{\overline{\cal S}}}\left(\rho_{i j}, \alpha_s \right)
\,+\, \Delta^{{\overline{\cal S}}} \left(\rho_{i j}, \alpha_s \right)\,,
\label{delta}
\end{equation}
where the correction term $\Delta^{{\overline{\cal S}}}$ must satisfy
\begin{equation}
\sum_{j \neq i} \frac{\partial}{\partial \ln(\rho_{i j})} \,
\Delta^{{\overline{\cal S}}} \left( 
\rho_{i j}, \as \right)  =  0 \qquad \Leftrightarrow \qquad 
\Delta^{{\overline{\cal S}}}  = \Delta^{{\overline{\cal S}}} 
\left( \rho_{ijkl}, \as \right) \, .
\label{deltaeq}
\end{equation}
In words, corrections to the dipole formula must be functions 
of the conformal cross rations $\rho_{ijkl}$: thus, by
eikonal exponentiation, they must arise from gluon webs 
connecting at least four hard partons, and therefore they can 
first appear at three loops. This analysis explains the results of 
Ref.~\cite{Aybat:2006wq}: indeed, the dipole formula correctly reproduces all existing finite-order results.

One may proceed to ask whether further constraints are available,
going beyond Eq.~(\ref{constraint}), that might force the
correction term $\Delta^{\overline{\cal S}}$ to vanish, or 
determine its color structure, and its functional dependence on
$\rho_{ijkl}$'s. This analysis, started in Ref.~\cite{Becher:2009cu}, 
was pursued in Ref.~\cite{Dixon:2009ur}. 

A general constraint on $\Delta^{\overline{\cal S}}$ is dictated 
by its behavior in the limit where two or more of the hard partons
become collinear. In this limit Eq.~(\ref{facto}) breaks down, 
however it is expected that the new singularities that arise 
should be captured by a splitting matrix depending only on 
the degrees of freedom of the partons becoming collinear. 
As shown in Refs.~\cite{Becher:2009cu,Dixon:2009ur}, this 
essentially forces $\Delta^{\overline{\cal S}}$ to have trivial 
collinear limits. One may furthermore impose Bose symmetry 
(since $\Delta^{\overline{\cal S}}$ arises diagrammatically from 
webs of gluons), and require that, at any given order, the functions 
comprising $\Delta^{\overline{\cal S}}$ should satisfy a 
transcendentality bound (at $g$ loops, one needs $\tau_{\rm max} 
= 2 g  - 1$). Remarkably, the set of functions satisfying all the 
constraints is quite small, though not empty. As an example, if one 
considers functions built out of products of logarithms of 
$\rho_{ijkl}$'s (which would naturally occur in Feynman diagram 
calculations), then Bose symmetry and collinear consistency single out 
a unique class of functions. Defining $L_{ijkl} \equiv \log \rho_{ijkl}$, 
the quadrupole component of $\Delta^{\overline{\cal S}}$, which 
is the basic building block for higher-point corrections as well, must 
be of the form
\begin{equation}
\hspace{-2pt}
\Delta_4 (\rho_{ijkl}) \, = \, 
\, {\bf T}_1^{a} {\bf T}_2^{b} {\bf T}_3^{c} {\bf T}_4^{d}
\left[ \, f_{ade}^{\phantom{ade}} \, f_{cb}^{\phantom{cb} e} \,  
L_{1234}^{h_1} \, \left(
L_{1423}^{h_2} \, L_{1342}^{h_3} \, - \, (-1)^{h_1 + h_2 + h_3} \, 
L_{1342}^{h_2} \, L_{1423}^{h_3} \right) \, + \, {\rm cycl.} \right],
\label{delta4log}
\end{equation}
where $h_i$ are positive integers, and only cyclic permutations of the 
$(2,3,4)$ labels must be added. Including the transcendentality 
bound, at three loops precisely one function in this class 
survives~\cite{Dixon:2009ur}: it is given by Eq.~(\ref{delta4log})
with $h_1 = 1$ and $h_2 = h_3 = 2$. If one 
considers more general classes of functions, including for example 
polylogarithms, at least two further consistent examples can be found. 

The question whether the dipole formula receives corrections involving 
multigluon correlations at high perturbative orders remains thus open. 
It is clear however that such corrections, if any, are constrained in a
much stronger way than might have been expected, and one may 
concretely hope to bring them under control in the not too distant 
future.


\begin{thebibliography}{99}

\bibitem{reviews}
  J.~C.~Collins,
  Adv.\ Ser.\ Direct.\ High Energy Phys.\  {\bf 5} (1989) 573
  [hep-ph/0312336];
  G.~Sterman, hep-ph/9606312;
  N.~Kidonakis, G.~Oderda and G.~Sterman,
  Nucl.\ Phys.\  B {\bf 531} (1998) 365
  [hep-ph/9803241];
  L.~Magnea, Pramana {\bf 72} (2008) 1
  [arXiv:0806.3353 [hep-ph]].

\bibitem{Magnea:1990zb}
  L.~Magnea and G.~Sterman,
  Phys.\ Rev.\  D {\bf 42} (1990) 4222.

\bibitem{Dixon:2008gr}
  L.~J.~Dixon, L.~Magnea and G.~Sterman,
  JHEP {\bf 0808} (2008) 022
  [arXiv:0805.3515 [hep-ph]].

\bibitem{Korchemsky:1985xj}
  G.~P.~Korchemsky and A.~V.~Radyushkin,
  Phys.\ Lett.\  B {\bf 171} (1986) 459.

\bibitem{Alday:2009zf}
  L.~F.~Alday,
  JHEP {\bf 0907} (2009) 047
  [arXiv:0904.3983 [hep-th]].

\bibitem{Sterman:2002qn}
  G.~Sterman and M.~E.~Tejeda-Yeomans,
  Phys.\ Lett.\  B {\bf 552} (2003) 48
  [hep-ph/0210130].

\bibitem{Bern:2005iz}
  Z.~Bern, L.~J.~Dixon and V.~A.~Smirnov,
  Phys.\ Rev.\  D {\bf 72} (2005) 085001
  [hep-th/0505205].

\bibitem{Gardi:2009qi}
  E.~Gardi and L.~Magnea,
  JHEP {\bf 0903} (2009) 079
  [arXiv:0901.1091 [hep-ph]].

\bibitem{Becher:2009cu}
  T.~Becher and M.~Neubert,
  Phys.\ Rev.\ Lett.\  {\bf 102} (2009) 162001
  [arXiv:0901.0722 [hep-ph]];
  JHEP {\bf 0906} (2009) 081
  [arXiv:0903.1126 [hep-ph]].

\bibitem{Gardi:2009zv}
  E.~Gardi and L.~Magnea,
  arXiv:0908.3273 [hep-ph].

\bibitem{Aybat:2006wq}
  S.~M.~Aybat, L.~J.~Dixon and G.~Sterman,
  Phys.\ Rev.\ Lett.\  {\bf 97} (2006) 072001 
  [hep-ph/0606254];
  Phys.\ Rev.\  D {\bf 74} (2006) 074004
  [hep-ph/0607309];
  L.~J.~Dixon,
  Phys.\ Rev.\  D {\bf 79}, 091501 (2009)
  [arXiv:0901.3414 [hep-ph]].

\bibitem{Dixon:2009ur}
  L.~J.~Dixon, E.~Gardi and L.~Magnea,
  arXiv:0910.3653 [hep-ph].

\end{thebibliography}
\end{document}